\documentclass[letterpaper,pra,twocolumn,showpacs,superscriptaddress]{revtex4}

\usepackage{graphicx}
\usepackage{amsmath}
\usepackage{amssymb}

\newcommand{\bra}[1]{\langle#1|}
\newcommand{\ket}[1]{|#1\rangle}

\begin{document}
\title{Continuous Variable Quantum Information Processing}

\author{Ulrik L. Andersen}
\affiliation{Department of Physics, Technical University of Denmark, Building 309, 2800 Lyngby, Denmark}
\author{Gerd Leuchs}
\affiliation{Max Planck Institute for the Science of Light, G\"unther Scharowsky Strasse 1, 91058 Erlangen, Germany}
\affiliation{University Erlangen-N\"urnberg, Staudtstrasse 7/B2, 91058 Erlangen, Germany}
\author{Christine Silberhorn}
\affiliation{Max Planck Institute for the Science of Light, G\"unther Scharowsky Strasse 1, 91058 Erlangen, Germany}

\begin{abstract}
Observables of quantum systems can posses either a discrete or a continuous spectrum. For example, upon measurements of the photon number of a light state, discrete outcomes will result whereas measurements of the light's quadrature amplitudes result in continuous outcomes. If one uses the continuous degree of freedom of a quantum system either for encoding, processing or detecting information, one enters the field of continuous variable (CV) quantum information processing. In this paper we review the basic principles of CV quantum information processing with main focus on recent developments in the field. We will be addressing the three main stages of a quantum informational system; the preparation stage where quantum information is encoded into CVs of coherent states and single photon states, the processing stage where CV information is manipulated to carry out a specified protocol and a detection stage where CV information is measured using homodyne detection or photon counting.
\end{abstract}

\maketitle

\section{Introduction}
One of the most central and counter-intuitive postulates of quantum mechanics is the superposition principle which states that two or more physical states of the same system can co-exist. This simple postulate has led to very accurate descriptions of physical phenomena that are not explainable by classical mechanics. At first the controlled operation of individual quantum systems was used to put quantum theory to more and more stringent tests. Examples are the measurements of Einstein-Podolsky-Rosen entanglement~\cite{Reid2008} and the violation of the Bell-inequalities~\cite{Aspect1982}, both of which are closely related to the superposition principle. Today the mastering of individual quantum systems, their controlled operation and interplay are the basis of quantum information processing and communication. This development is giving rise to the emerging field of quantum engineering. In quantum computation and quantum communication, the basic laws of quantum mechanics are now used to enable communication with absolute security and, for the future, they offer the potential of very fast execution of complex computational algorithms~\cite{NielsenBook}. While quantum computers will threaten the security of traditional communication, and while the presently used classical encryption schemes are anyway not proved to be secure, quantum techniques can on the other hand help to regain security. In quantum cryptography the encryption problem reduces to key distribution, or more precisely to key growing in combination with authentication~\cite{Dusek2006}.

From a theoretical point of view there are different ways to describe a quantum system: with discrete or  continuous variables~\cite{loock}. In this review we will focus on the latter. The distinction between discrete and continuous variables is related to the effect quantization has in this two cases. For light the quantization of the field does not change the spatial-spectral mode pattern but rather the possible excitation per mode. If we consider the energy we will find a discrete spectrum, but if we look at field variables such as amplitude or phase, their spectrum is continuous and quantization manifests itself in unavoidable quantum uncertainties as described by Heisenberg's uncertainty relation. It is worth noting that the uncertainty relation also applies to the discrete variable case if the system is in a superposition of two or more energy eigen states. The distinction is that in the discrete case the commutator leading to a Heisenberg uncertainty is operator valued~\cite{Korolkova2002}. If the expectation value of this operator is zero the resulting uncertainty product has no lower bound, unlike in the case of the continuous field variables.

For light fields any pure or mixed quantum state can be described in the photon number basis, but it is  always possible to convert this description to the continuous variable language and vice versa~\cite{SchleichBook}. This is discussed in more detail in the next section, but we would like to emphasize at this point, that the choice of a preferable representation, either with discrete or continuous variables, ultimately depends on the experiment to be described. If we use photon number resolving detectors the photon number basis expansion is an obvious choice. The same is true for dichotomic 'click' detectors which have only two possible outputs, reporting either 'no photon' or 'one or more photons'. In contrast, for homodyne detection relying on the interference of the signal field with a local oscillator the continuous variable approach is the better alternative. Discrete and continuous variables thus do not correspond to either few or many photons, both rather apply to both regimes of photon numbers. In practice this symmetry is broken because the higher the excitation of a mode the higher the technological challenge for the experimenter to build a detector still able to distinguish different photon numbers. Therefore the details of a discrete variable description of an intense light field are essentially not accessible experimentally. A particularly intriguing situation appears if one combines both types of detectors in one set-up, e.g. conditioning the homodyne measurement on one mode on the 'click' detection of another necessarily weakly excited mode. In this way it is possible to transform a Gaussian squeezed state into a non Gaussian state~\cite{dakna1997.pra} (sec. 4.2)

\begin{figure}[htb]
  \includegraphics[width=\linewidth]{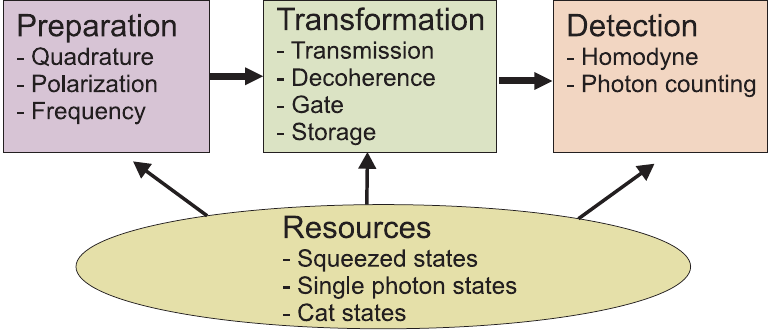}
  \caption{General setup for any quantum information protocol.}
  \label{GeneralSetup}
\end{figure}

The field of quantum information processing can be roughly divided into three subfields: Quantum key distribution, quantum computation and quantum information distribution including quantum memories. The continuous variable version of these protocols will be addressed in this review. However, before jumping into the details of the protocols we go through the various ingredients needed to build up the complex networks. Any quantum information protocol or any quantum mechanical optical experiment consists of three parts. In the first part, the optical quantum information is prepared by encoding information into an optical quantum state such as coherent state of light or a single photon. The preparation stage is followed by an operation which either cannot be controlled by the experimenter, i.e. the state will decohere as a result of a coupling to the environment, or the operation can be under full control and a dedicated quantum protocol is executed onto the state. Finally, the last part of the experiment is a measurement where the resulting optical state is either characterized by performing an ensemble measurement or information is decoded from the state through a single shot measurement. A schematic of such a general setup is shown in figure~\ref{GeneralSetup}. The review paper will be structured according to such a scheme. We will review various ways of preparing continuous variable quantum information (sec. 2), detecting CV quantum states (sec. 3) and operating onto the states (sec. 4). For all three parts, non-classical resource states such as single photon states, squeezed states, entangled states and cat states are important. The generation of such states will therefore also be reviewed (sec. 5). With all these tools at hand we finalize the article by reviewing some of the most important CV quantum information protocols such as quantum key distribution, quantum memory, quantum computation and quantum information distribution (sec. 6).     

\section{Quantum information}

In most quantum information protocols, the standard unit of information is the quantum bit (also known as the qubit). Information is carried as a superposition of two orthogonal, pure quantum states: $|\phi\rangle=\alpha |0\rangle+ \beta |1\rangle$ where $|\alpha |^2+|\beta |^2=1$. This definition implicitly assumes that the information carrier itself constitutes a single particle, i.e. in the case of optical fields qubits are typically associated with single photon states.  Information coding is accomplished by using modal properties, e.g. the polarization of the photon, and employs two or a finite number of basis states. For the extension to higher dimension the use of superposition states involving three or four distinguishable single photons have been introduced to allow for qutrits or qutrats, respectively. Still, all discrete variable systems have in common that they are designed for state preparation and information read-out, which is based on Fock states and photon counting. However, both of these requirements are experimentally challenging and become less and less feasible if states with increasing photon numbers or networks with multiple input states are considered.

As an alternative approach  CV encoding has been introduced, which is adapted to the use of laser light and employs homodyne measurements instead of single photon detection. Homodyne detection yields information about the field quadratures of a quantum state, which are related to amplitude and phase properties of the light. In the quantum mechanical description quadratures correspond to the position and momentum of a local oscillator and constitute an infinite dimensional Hilbert space. Quantum information can then be defined as a continuous superposition of eigenstates; $|\xi\rangle\propto\int \langle x|\phi\rangle |x\rangle dx$ where $\langle x|\phi\rangle$ is the wave function of the state in the continuous basis $x$~\cite{SchleichBook}. A common basis for continuous variables quantum information is the position eigenbasis (or equivalently, the momentum eigenbasis). Besides being associated with the position degree of freedom of a particle, the position state is also an eigenstate of the amplitude quadrature of a light mode and a phonon mode (e.g. the vibrational modes of an atom). Thus CV quantum information can be carried by any light and atomic system. 

\begin{figure}[htb]
  \includegraphics[width=\linewidth]{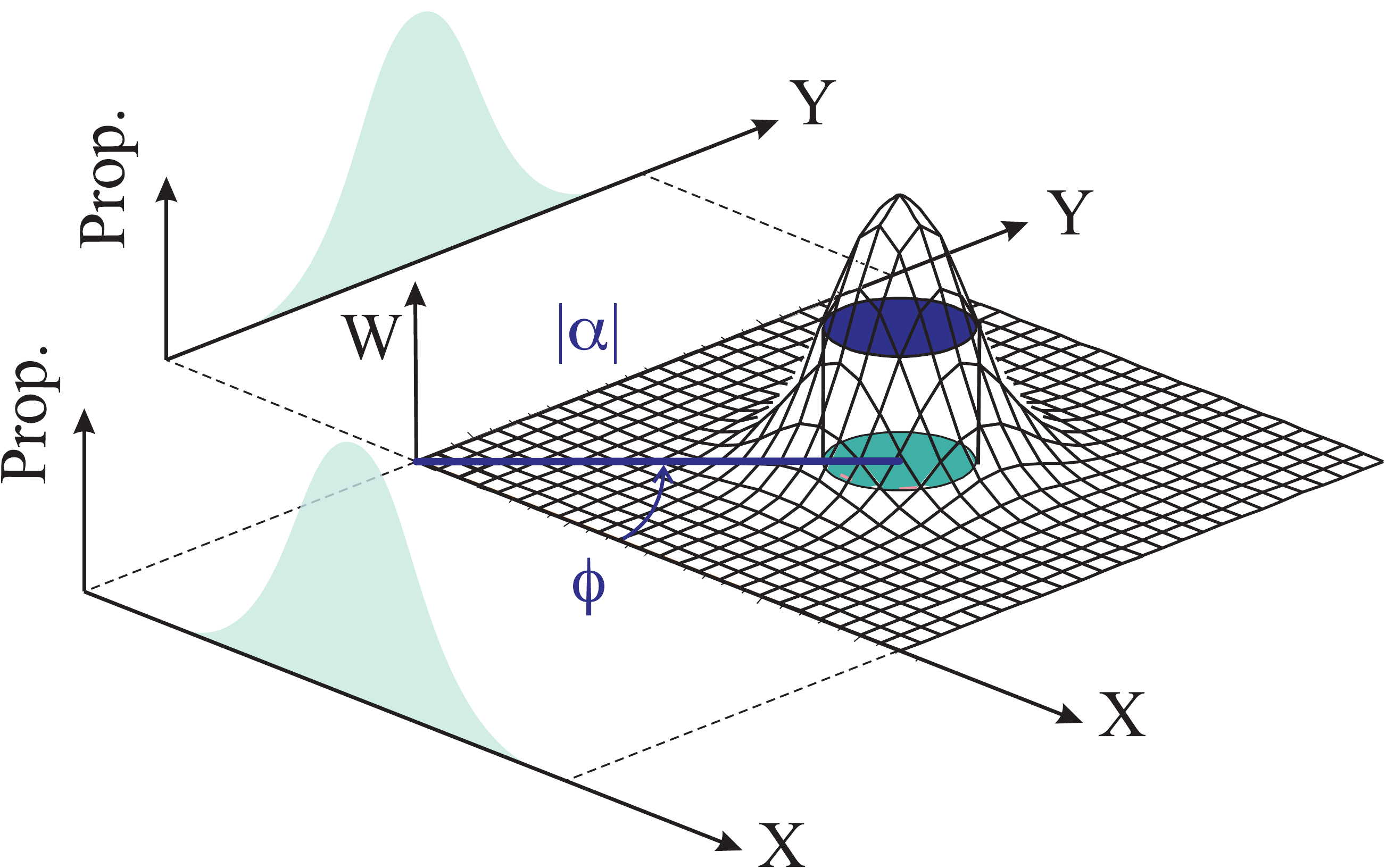}
  \caption{ (from  \cite{silberhorn2007}) Wigner function of a coherent state in phase space.  Integrating over quadrature  $X$ ($Y$) yields the probability distribution for $Y$ ($X$)}
  \label{Wigner}
\end{figure}

The genuine quantum character of encoded CV information also manifests itself in the Heisenberg uncertainty of the conjugate amplitude and phase quadrature observables, or the orthogonal quadratures $X$ and $Y$ of light fields. The first CV protocols were actually designed to exploit the uncertainty relationships directly for quantum information applications. For that the properties of photonic states can be visualized in a phase space representation, where the axis are defined by a pair of orthogonal quadratures and a classical optical field would be determined by a single point corresponding to its complex field amplitude. However, for quantum states of light the definition of  a single point in phase space is problematic, because conjugate variables cannot be measured simultaneously with arbitrary precision. Thus, we have to introduce a Wigner-function, which serves as a quasi-probability distribution in phase space and provides a direct link to homodyne detection used for observing quadratures. The probability distribution of a quadrature measurement $X$ is obtained from the Wigner function by integration over the conjugate quadrature $Y$ (see Fig.~\ref{Wigner}).

Gaussian states are defined as quantum states, which exhibit a Wigner function with Gaussian marginal distributions. Gaussian states are thus fully characterized by the mean and variance values of the extreme marginal distributions, or the first and second order moments respectively. Traditionally CV quantum communication was solely based on Gaussian states and many protocols are formulated in terms of uncertainty variances neglecting higher order moments. Laser light as coherent states corresponds to a minimum uncertainty state with a Gaussian Wigner function and a symmetric distribution of the quadrature components, i.e.
$\bra{\alpha} \Delta^2 \hat{X}_\phi \ket{\alpha} = \bra{\alpha} \Delta^2
\hat{X}_{\phi'} \ket{\alpha}  $
for all angles $\phi,\phi'$ with generalized quadratures $\hat{X}_\phi = \hat{a} e^{-i\phi} + \hat{a}^{\dag} e^{i\phi}$. More formally a coherent state with field amplitude $\alpha= \alpha_r+\alpha_i$  is defined by
$\ket{\alpha}= \exp(\alpha \hat{a}^{\dag} -\alpha^{*} \hat{a} ) \ket{0},$
and its 
 Wigner function  reads \[W(X,Y)=\frac{2}{\pi}\, \exp\left(-2[(X-\alpha_r)^2+(Y-\alpha_i)^2]\right).\]   
Squeezed states also exhibit Gaussian Wigner functions, but the uncertainty of one quadrature is reduced at the expense of an increase of the conjugate one: $\bra{\zeta} \Delta^2 \hat{X}_\phi \ket{\zeta} < \bra{\alpha} \Delta^2\hat{X}_{\phi} \ket{\alpha}  $ and $\bra{\zeta} \Delta^2 \hat{X}_{\phi+\frac{\pi}{2}} \ket{\zeta} > \bra{\alpha} \Delta^2\hat{X}_{\phi+\frac{\pi}{2}} \ket{\alpha}$. Restricting oneself to pure vacuum squeezed states they can be defined by $\ket{\zeta}= \exp(-\frac{1}{2}\zeta (\hat{a}^{\dag})^2 + \frac{1}{2}\zeta^{*} \hat{a}^2 ) \ket{0}. $ In this description the squeezed quadrature is oriented along $\varphi$ and the squeezing can be quantified by $\bra{\zeta} \Delta^2 \hat{X}_\varphi \ket{\zeta} = e^{-2r}\bra{\alpha} \Delta^2\hat{X}_{\phi} \ket{\alpha}$. For $\varphi=0 $ the respective Wigner functhin is then given by
 \[W(X,Y)=\frac{2}{\pi}\, \exp\left(-2[e^{2r}X^2+e^{-2r}Y^2]\right);\]  rotations and displacements in phase space can be included by the corresponding tranformations of the quadratures $X$ and $Y$. Note, that that  extreme squeezing values $r \rightarrow \infty $ with $\bra{\zeta} \Delta^2 \hat{X}_\varphi \ket{\zeta} \rightarrow 0$ also entail that  $\bra{\zeta} \Delta^2 \hat{X}_{\varphi+\frac{\pi}{2}} \ket{\zeta} \rightarrow \infty$, which, in turn, implies an infinite energy content. Therefore extreme squeezing values are unrealsitic and $\bra{\zeta} \Delta^2 \hat{X}_\varphi \ket{\zeta} = 0$ is not possible even in principle. For mixed squeezed states the anti-squeezed quadrature possess a noise variance, which is higher than  required by the Heisenberg uncertainty product and the Wigner functions have to be modified accordingly with increased variances for the Gaussian functions.

Recent progress in CV quantum information now goes beyond Gaussian states and investigates the employment of non-Gaussian states for advanced systems. One main driving force behind this development has been provided by the recognition that Gaussian states in combination with Gaussian operation are not sufficient to achieve entanglement distillation  or CV quantum computation  (see Sec.\ref{sec:entdist}). This has led to the exploration of various approaches to implement non-Gaussian states and to study hybrid systems using homodyne detection and photon counting in one single system.

In this context it is important to note that most quantum systems have both, finite and continuous degrees of freedom. In other words the complete description of a given quantum states has to take into account all degrees of freedom of an optical field, i.e. its photon number, wavelength, polarization, and wavevector. To illustrate the interplay of discrete and continuous variables we might consider a single photon state as the simplest example. The photon number is a discrete variable, but quadrature measurements yield CV outcomes, the polarization degree of freedom is always binary, but the spectral and spatial degree of freedoms can be discrete and continuous. A single photon (as well as many other states) can therefore be used as information carrier both for discrete variable and continuous variable quantum information processing. Since different degrees of freedoms are often coupled due to the state generation process, e.g. photon pair generation always results in spatio-spectral correlations due to energy and momentum conservation, a description with decoupled degrees of freedom requires the introduction of a multi-mode structure of the states \cite{law2000,Opatrny2002,Valcarcel2006}.  A problem arises if we want to combine conditioning photon counting with CV states, because the modal properties of detector response frequently cannot be matched with the internal structure of the CV quantum states, which, in turn, results in mixed states \cite{Rohde2007}. On the contrary, if different degrees of freedom can be decoupled, i.e. if their eigenstates form a tensor product, the individual quantum systems can be treated independently, and the actual system being considered depends only on the observable that is monitored by the detector in the physical setup. Current efforts focus on state engineering including all degrees of freedom to allow for optimized state preparation for different purposes, such as maximizing the information transfer or the efficient conditional preparation of pure states \cite{Uren2005,Mosley2008}.

A second motivation to investigate non-Gaussian states more deeply comes from a different approach to read out quantum information of a CV system \cite{Ralph2003}. In analogy to qubit system one can also realize the basis states  $|0\rangle$  and  $|1\rangle$ as two coherent states with different classical amplitudes, e.g. $|\alpha\rangle$ and $|-\alpha\rangle$ , which correspond to the respective displacements in phase space. Now, defining a qubit in this implementation requires the coherent superposition of  $|\alpha\rangle$ and $|-\alpha\rangle$ and yields, for example, the states $|\phi_+\rangle\propto|\alpha\rangle+|-\alpha\rangle$ or $|\phi_-\rangle\propto|\alpha\rangle-|-\alpha\rangle$. Though being composed of two Gaussian states, the states $|\phi_+\rangle $ and  $ | \phi_- \rangle $ themselves are highly non-Gaussian and, in fact, constitute Schr\"odinger cat states for large values of field amplitudes. This type of CV encoding is much more intricate and necessitates complex state preparation, but it offers the benefit that all required couplings between different channels in quantum networks are linear and can be set up with phase shifters and beam splitters.

\section{Quantum detection}

Measuring quantum properties of photonic states always means the detection of photon or field statistics of some type, but the way we interpret our results can be quite different and depends ultimately on the type of measurement and on the different experimental imperfections, which limit the resolution and/or the noise figures of the recorded data. 

For intense light beams we can use standard photodiodes and study quantum characteristics by direct detection. Standard photo-diodes can have a high quantum efficiency and thus are well suited for an accurate mapping of the field statistics onto detected photocurrents. Moreover, the detector and amplifier noise is typically much lower than the signal and can be largely ignored. However, measurements with usual photo-diodes have limited resolution, which means that they are not capable to distinguish between individual photon numbers. Therefore it is customary to decompose the field operator â for large field amplitudes into a "classical" displacement $\alpha$ and the quantum operator $\delta\hat a$, such that $\hat a=\alpha+\delta\hat{a}$. Considering only the first order terms of $\delta\hat a$ \cite{Fabre1990} we obtain for a photon number measurement the relationship $\hat N=\alpha^2+\alpha\delta \hat X$, where we have defined the amplitude quadrature operator as $\delta\hat X=\delta\hat{a}+\delta\hat{a}^\dagger$. This shows that in the limit of sufficiently large classical field amplitudes the generated photocurrents yield direct information about the amplitude quadrature, while the uncertainty of the orthogonal phase quadrature $\delta\hat Y=i(\delta\hat{a}-\delta\hat{a}^\dagger)$ cannot be seen.

 Direct detection offers the advantage that the bright beam "carries" its own local oscillator and thus intrinsically determines a fixed phase reference.  For practical quantum communication systems this also implies that the communicating parties do not have to synchronize their measurement bases before data transmission, but the bright quantum states themselves allow them to establish a shared reference frame.  However, all CV quantum protocols as well as the characterization of a Wigner function require the observation of a pair of conjugate variables; but direct detection lacks any information about the phase quadrature. To circumvent this drawback the use of Stokes operators has been suggested \cite{Korolkova2002}. The Stokes operators describe the polarization properties of quantum fields and can all be measured by direct detection in conjunction with appropriate wave plates and polarizing beam splitters. In the quantum mechanical description the Stokes operators satisfy the commutation relations of the su(2) algebra, which means that a suitable pair of conjugate Stokes observables fulfilling an operator valued Heisenberg uncertainty relationship can be defined.  Several experiments have demonstrated the utility of this approach in the context of CV quantum information processing with intense fields \cite{Heersink2003,Bowen2002}, and a tomographic reconstruction of the Wigner function of a polarization squeezed field has been accomplished by Marquardt et al.\cite{Marquardt2007} in 2007.

For dark quantum states, i.e. $\alpha\approx 0$, direct detection of any signal necessitates either the use of detectors with single-photon sensitivity, or the signal state has to be "amplified" by the interference with a separate strong local oscillator reference beam and subsequently detected with standard diodes. In the former case a photon number resolution of $\delta n\approx 100$ may be enough to perform meaningful experiments \cite{Gottesman2001}. The latter setup constitutes the well-established homodyne detection method, where the relative phase between the signal and the local oscillator determines which generalized quadrature component $\hat X^\phi=\hat a e^{-i\phi}+\hat a^\dagger e^{i\phi}$ is measured. By choosing several quadrature angles $\phi$  a tomographic reconstruction of the corresponding Wigner function becomes possible, which uniquely defines the quantum state (see Fig.~\ref{Wigner}). This technique has first been introduced by Smithey et al. \cite{Smithey1993} in 1993, and since then has become the workhorse for many CV systems.  More recently it has been applied not only to Gaussian states, but in 2001 Lvovsky et al. \cite{lvovsky2001.prl} employed it for the first time to characterize a non-Gaussian state, namely a single-photon Fock state with a negative Wigner function. In the last years a few groups have demonstrated homodyne tomography for Schr\"odinger cat like states \cite{ourjoumtsev2006.sci,nielsen2006.prl,wakui2007.oex}. 

\begin{figure}[htb]
 \includegraphics[width=0.95\linewidth]{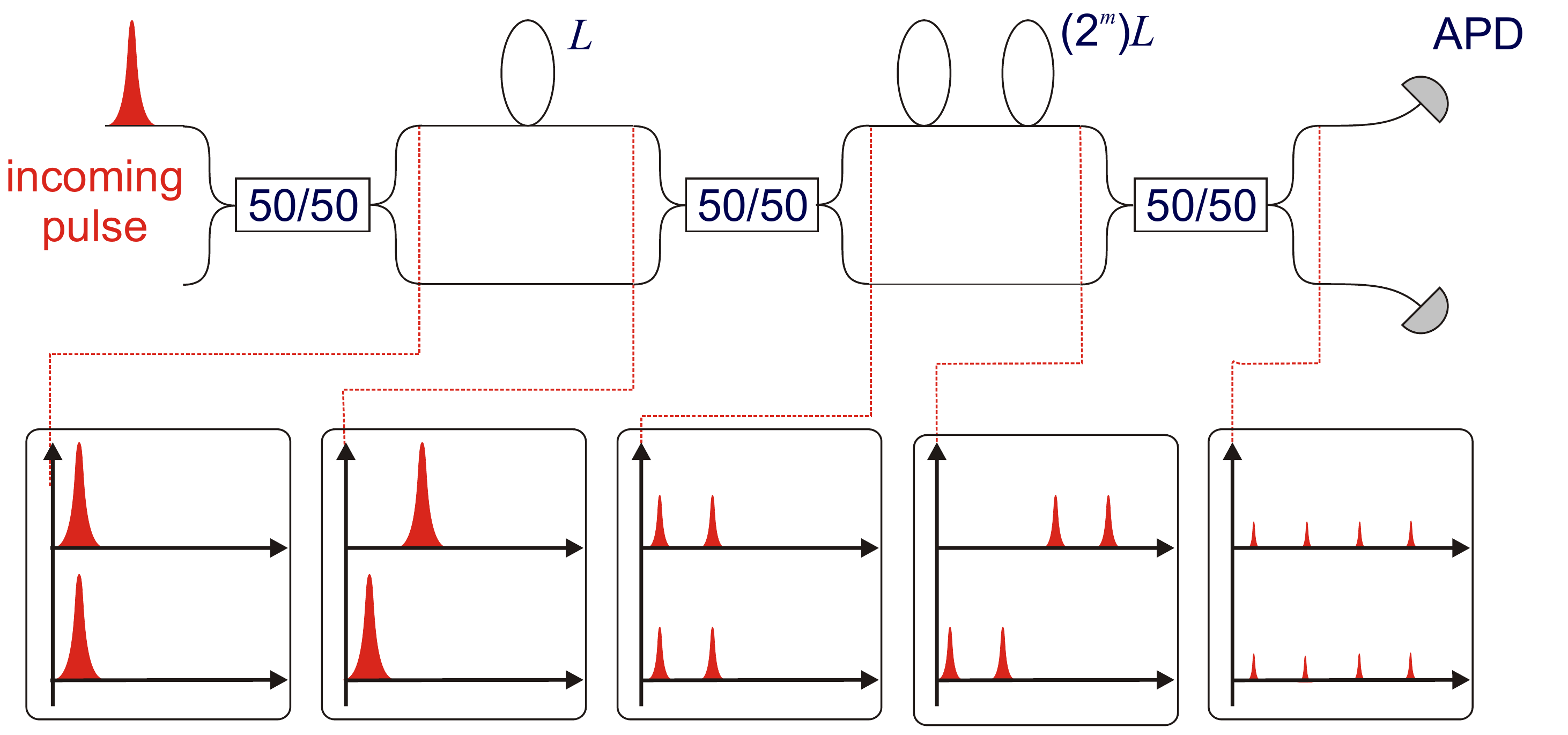}
  \caption{(from \cite{silberhorn2007}) Time-Multiplexed Detector (TMD): An input pulse is divided by means of a fiber optical network with two spatial channels and adapted fiber delay lines into several output pulses; subsequent detection with APDs allows to measure the photon number of the impinging pulse}
  \label{TMD}
\end{figure}

An alternative way of characterizing multi-photon states with low mean photon number can be provided by direct detection, if a photon number resolving detector is available. Such methods become especially relevant in the context of non-Gaussian states, where performing conditioning state preparation using photon counters is essential. Various efforts have been made in this direction over the last years, including superconducting devices, visible light counters or detectors employing an avalanche process, which is not driven in the Geiger mode (for review, see e.g. \cite{silberhorn2007}). A different way to achieve photon number resolution by optical means can be realized by multiplexed detection in combination with standard binary avalanche photo diodes (APD). A time-multiplexed detector (TMD) utilizes two spatial modes and a selectable number of temporal modes \cite{Achilles2003,Fitch2003}. A fiber network with, e.g. three, 50/50 couplers and fiber loops of variable lengths (see Fig. \ref{TMD}) serves as a multiplexing device, dividing the input pulse into eight  output pulses, which can be measured with two APDs. The resulting "click" statistics provides information about the impinging photon statistics, which can even be precisely specified by detector tomography \cite{Lundeen2008}. In the standard TMD modeling two experimental imperfections, namely the finite probability of two photons not being separated by the fiber network and the overall losses in the system, are taken into account. For ensemble measurements these deficiencies can be overcome if a precise loss calibration is possible, which can provide a loss-tolerant characterization of the photon statistics \cite{Achilles2006}. It is interesting to note that in contrast to standard homodyne detection, a TMD is sensitive to all modes of an input state, because no interference with a local oscillator is needed. This enables, e.g. the identification of multi-mode characteristics of twin beams, which is not observable with standard homodyne measurements \cite{Avenhaus2008}. 

Nevertheless for the complete characterization of quantum light, $|\Psi\rangle=\sum_n c_n | n\rangle$, the knowledge of photon statistics is not sufficient, because the phase relationships between the different photon number contributions are not determined. However, the value of the Wigner function of a state at the origin is directly related to the photon statistics $p_n= |c_n|^2$ by the parity of the state $W(0)=\sum_n(-1)^np_n$, which can provide us with some intuition how to determine the Wigner function by photon counting. Wallentowitz and Vogel \cite{Wallentowitz1996}
 and Banaszek et al. \cite{Banaszek1996} have proposed to measure the Wigner function by parity measurements in conjunction with a series of appropriate displacements in phase space. Experimentally, this can be accomplished in a noisefree way by superimposing the signal field with  strong reference beam at a highly asymmetric beam splitter. This technique corresponds to probing the Wigner function point by point and could allow for studying interesting regions in phase space, e.g. the negative parts, independent from other areas. While this characterization method has not yet been accomplished for traveling light fields, the principle has been demonstrated for a field stored in a high-Q cavity \cite{Bertet2002}.

\section{Quantum operations}

Quantum operations can be either Gaussian or non-Gaussian. The Gaussian operations are those that map a Gaussian state onto another Gaussian state whereas non-Gaussian operations map a Gaussian state onto a non-Gaussian state. We now discuss these two types of transformation.

\subsection{Gaussian operations}
\begin{figure}     
   \includegraphics[width=0.9\columnwidth]{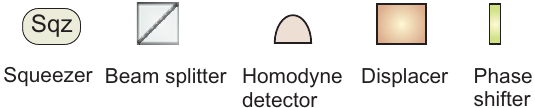}
 \caption{Basic optical components that transform a Gaussian state onto another Gaussian state. The Hamiltonian of these components are at most quadratic in the quadratures.}
 \label{Gaussian_optics}
 \end{figure}

\begin{figure}
       \includegraphics[width=0.8\columnwidth]{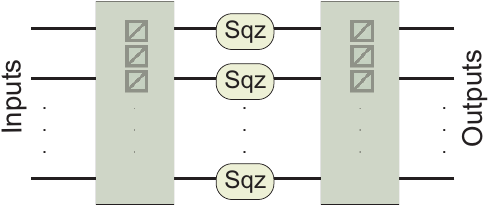}
 \caption{Illustration of Bloch-Messiahs reduction theorem. It states that any multi-mode Gaussian transformation can be accomplished by inserting single mode squeezers in each of the arms of an interferometer.}
 \label{B-M}
 \end{figure}

%%% homodyne detection is missing !!!!!!!! and in fig4 the phse shifter is missing
The elementary and unitary Gaussian transformations are the beam splitting, the phase shifting, the displacing, the squeezing operation and the homodyne detector, see Fig.~\ref{Gaussian_optics}. By combining these operations in an optical circuit, an arbitrary Gaussian transformation can be implemented. In fact, it has been shown  by using the so-called Bloch-Messiahs decomposition theorem, that an arbitrary multi-mode Gaussian transformation can be implemented with an array of beam splitters followed by an array of single mode squeezers and another array of beam splitters as illustrated in Fig.~\ref{B-M}~\cite{braunstein2005.pra}. This is an important result since it says that any complex multi-mode Gaussian transformation can be simply executed by utilizing single mode squeezers in a multi-mode interferometer. The challenge in implementing an arbitrary Gaussian operation therefore reduces to the challenge in implementing a pure and efficient single-mode squeezing operation.

\begin{figure}
     \includegraphics[width=0.8\columnwidth]{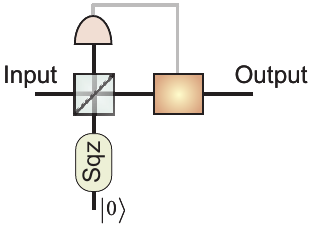}
 \caption{Schematic of the setup for an optical squeezing transformation using off-line prepared squeezed vacuum. The squeezed vacuum merges with a coherent state on a beam splitter with a variable beam splitting ratio, and the anti-squeezed quadrature of one output is measured. Finally, the continuous outcomes are used to drive a phase modulator traversed by the signal beam, thus implementing the displacement operation.}
 \label{sqz_ope}
 \end{figure}

Although remarkable progress has been made in developing new and improved techniques for squeezing the vacuum state (see section 6), there has been very few experimental studies devoted to the squeezing of fragile quantum information. This is probably due to the difficulty in combating decoherence in the squeezing operation. A new method that largely circumvents decoherence has recently been put forward in ref.~\cite{filip2005.pra}. This scheme is based on electro-optic feed-forward where the quantum state under interrogation interferes with an off-line prepared squeezed vacuum state on a beam splitter, one output is measured using homodyne detection and the outcome is scaled and used to displace the second output of the beam splitter, see Fig.~\ref{sqz_ope}. Such a squeezing operation was experimentally investigated in ref.~\cite{yoshikawa2007.pra}. Another protocol based on a similar setup was experimentally analysed in ref.~\cite{lam1998.prl} and a comparison between the various feed-forward schemes has been discussed in ref.~\cite{andersen2009.book}.

 \begin{figure}   
   \includegraphics[width=0.8\columnwidth]{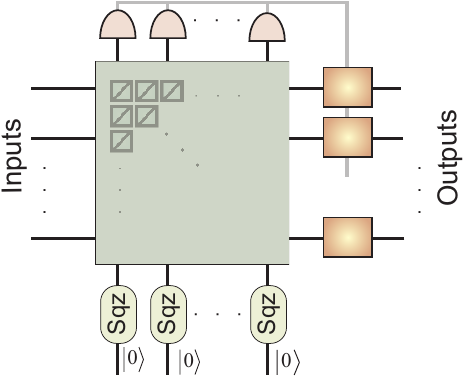}
 \caption{Uniting the feed-forward based squeezing transformation with Bloch-Messiah's reduction theorem, it is realized that any multi-mode Gaussian transformation can be accomplished by mixing the information carrying input states with off-line prepared squeezed vacua in a network of beam splitters followed by homodyne detection and feed-forward.}
 \label{BS}
 \end{figure}

 \begin{figure}
  \includegraphics[width=0.8\columnwidth]{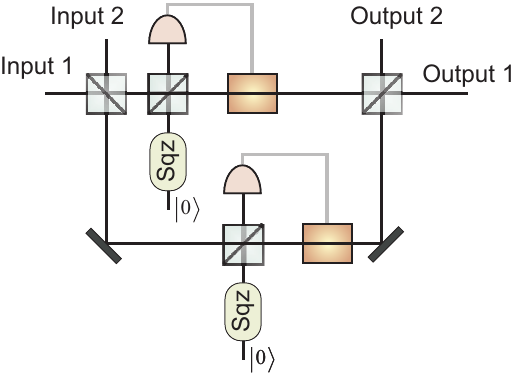}
 \caption{Illustration of a two-mode Gaussian transformation based on Bloch-Meassiah's reduction theorem and the feed-forward based squeezing operations. Two squeezers are placed in each of the arms of a Mach-Zehnder interferometer. This circuit has been used to implement a quantum nondemolition interaction.}
  \label{qnd}
  \end{figure}

 \begin{figure} 
    \includegraphics[width=0.8\columnwidth]{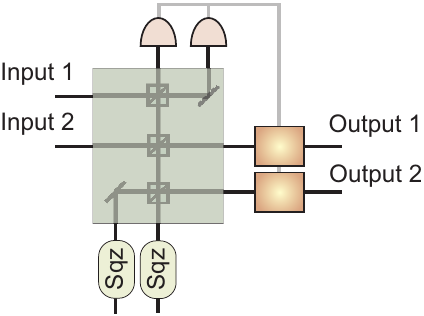}
  \caption{Optical circuit for an arbitrary two-mode Gaussian transformation. This is an alternative to the circuit in Fig.~\ref{qnd}. The interaction strength of the transformation can be controlled by the splitting ratio of the beam splitter marked by a star; the other beam splitters are symmetric. Without the squeezed vacuum resource, the circuit can still work as a linear amplifier at the quantum limit with a gain $G=1/T$ where $T$ is the transmittivity of the marked beam splitter.}
  \label{amp}
  \end{figure}

By combining the results of the Bloch-Messiahs decomposition theorem and the feed-forward based squeezed operation, we deduce that an arbitrary multi-mode Gaussian operation is realizable by coupling the input modes with squeezed vacuum modes in a beam splitter array, followed by measurements of some of the resulting output modes with homodyne detectors and displacement of the remaining modes. Arbitrary multi-mode Gaussian transformations can thus be implemented solely using linear optical transformations on the quantum state if the network is fed with squeezed vacua and the final operation is induced by homodyne measurements and displacements. This concept is illustrated in Fig.~\ref{BS}. It is interesting to note that this approach is reminiscent of the linear optical quantum computing approach suggested by Knill, LaFlamme and Milburn for qubits~\cite{knill2001.nat}, see Sec. 6. 

As a simple example we consider the arbitrary two-mode Gaussian operation. The setup is basically a Mach-Zehnder interferometer with a single mode squeezing operation placed in each of the two interferometer arms~\cite{filip2005.pra}. Using the feedforward based squeezing operations, the setup is composed of four simple linear beam splitting interactions, two squeezed vacuum resources, homodyne detection and feed-forward, see Fig.~\ref{qnd}. Using such a circuit, a two-mode Gaussian operation on two input coherent states were experimentally implemented in terms of the realization of a quantum non-demolition interaction~\cite{yoshikawa2008.prl}. The implemented circuit is identical to the optical parametric amplifier transformation where the two inputs and outputs of Fig.~\ref{qnd} correspond to the idler and signal modes of the amplifier. The fact that optical amplification can be enabled by linear optics, feedforward and off-line squeezed states was also suggested in ref.~\cite{josse2006.prl}. There, a different linear optical circuit was proposed as illustrated in Fig.~\ref{amp}. Interestingly, it was also realized and demonstrated that optical amplification of the signal state is possible at the shot noise limit without the use of the squeezed state resources~\cite{josse2006.prl}. This nonlinearity-free amplification transformation has been also employed to make coherent state clones at the quantum limit~\cite{andersen2005.prl}. Note that the coupling parameter (strength and phase) of the two-mode Gaussian transformation is controlled by the beam splitting ratio and the relative phase of the two incident modes.

An alternative approach to universal Gaussian transformations was recently proposed in ref.~\cite{menicucci2006.prl}. It was found that an arbitrary Gaussian transformation is implementable by employing an off-line prepared entangled Gaussian state (known as a cluster state) followed by homodyne measurements and feed-forward. It basically means that once the cluster state has been created, homodyne detection and feed-forward suffice to realize any multi-mode Gaussian transformation. A conceptual difference between the cluster approach and the feed-forward based squeezing approach is that in the former one the coupling parameters of the multi-mode interactions are solely controlled by the homodyne detectors as opposed to the latter one where the parameters are controlled by the beam splitting ratios. Therefore, the great advantage of the cluster state scheme is that the actual Gaussian coupling to be implemented can be decided upon detection, whereas in the feed-forward approach it must be decided before the linear beam splitter interaction. The main advantage of the feed-forward approach, however, is its physical simplicity compared to the cluster state approach. As an example illustrating this point, we consider the single mode squeezing operation. The feed-forward approach (discussed above) requires a vacuum squeezed state and a homodyne detector, whereas the cluster state approach for implementing a squeezing operation requires a five-mode entangled state and four homodyne detectors~\cite{loock2007.josa}.

\subsection{Non-Gaussian operations}

To implement an arbitrary pure operation, i.e. pure input states are mapped only onto arbitrary but pure output states, the Gaussian transformation must be complemented with a non-Gaussian one. Such an operation is, however, much harder to realize since it is associated with the introduction of a very large third order non-linearity. Sufficiently high non-linearities can in principle be achieved in a four wave mixing process in an optical fiber or in atomic vapor. Although the technology for implementing large non-linearities is improving, to date there has been no demonstrations of non-Gaussian transformations of Gaussian states using four wave mixing. There is, however, another approach to implement non-Gaussian transformations. This approach is based on measurement induced operations much like the Gaussian approaches introduced above. However, instead of using a homodyne detector, which transforms Gaussian state onto another Gaussian state, one uses highly non-linear detectors such as an avalanche photodiode (APD) which projects a Gaussian state into a non-Gaussian one. A simple example of such a transformation is the preparation of a single photon state from a weakly entangled two-mode Gaussian state (produced by e.g. a weakly pumped optical parametric amplifier): Upon detection of a single photon in one of the two modes (using an APD), the other mode is prepared in a highly non-Gaussian state, namely the single photon state. This transformation was beautifully demonstrated in an experiment by Lvovsky et al.~\cite{lvovsky2001.prl}. In a similar approach Zavatta et al. \cite{Zavatta2004}
employed  an optical parametric amplifier, which was stimulated by a weak coherent beam, in combination with a conditioning APD detection to generate 
 single-photon-added coherent states and to demonstrate the transition between particle like (i.e. non-Gaussian) and wave-like (i.e Gaussian) behavior. 
Other recent experiments demonstrating a measurement induced non-Gaussian transformation were carried out in the group of  Grangier~\cite{wenger2004.prl,ourjoumtsev2006.sci}. Following the proposal of ref.~\cite{dakna1997.pra}, they showed that a Gaussian squeezed state can be transformed into a superposition state of two coherent states by subtracting a single photon from a squeezed state~\cite{ourjoumtsev2006.sci}. This was done by reflecting a small part of the squeezed state at a beam splitter and detecting the presence of a photon with an APD. By conditionally selecting the remaining state based on the firing of the APD, a highly non-Gaussian state representing a small cat state was prepared and fully characterized using homodyne tomography. Photon subtraction has also been achieved by other groups~\cite{nielsen2006.prl,wakui2007.oex}, and new methods to produce cat states with larger excitations have been devised~\cite{ourjoumtsev2007.nat,takahashi2008.prl}. One main challenge for this type of experiments lies in the precise control of all degrees of freedom of the photonic states and detection setups, because APD detection typically features a different spectral-spatial mode response than homodyning. This can result in an unwanted multi-mode structure and mixedness for conditional state preparation.

The measurement induced non-Gaussian transformations have hitherto been used for state preparation (such as the single photon states and the cat states), but it can be also used to transform quantum states carrying information. For example, by complementing the homodyne detectors with photon number resolving detectors in the cluster state approach, it has been proposed that an arbitrary quantum operation can be realized. This will be discussed further in Sec.~\ref{computing}.

\section{Quantum resources}

Although all optical states are quantum states, it is customary to refer to a classical and a non-classical regime of states. The quantum-classical boundary is often defined through the behavior of the Glauber-Sudarshan P-function which is the probability function of the state in the diagonalized coherent state basis~\cite{KnightBook}. It is thus an alternative to the Wigner function. If the function is ill-behaved as a probability distribution thus exhibiting either singularities or negativities, the state is said to be non-classical. Thermal states which are describable as a mixture of coherent states have a well-behaved P-function whereas squeezed states give rise to an ill-behaved P-function, and therefore these two states are classified as being classical and non-classical, respectively. The P-function for a coherent state is a delta-function, thus it lies just on the quantum-classical boundary. This border between classical and quantum states is of high importance in terms of CV quantum information processing since it turns out that resource states that belong to the "classical" regime are incapable of executing quantum protocols beating the performance of classical protocols. Only when non-classical states are used, the superiority of quantum information processing is manifested. There is however an exception to that rule; it is possible to carry out the protocol of quantum key distribution even with states that are represented by a well-behaved P-function such as a Gaussian function provided that it is very narrow and thus close to a delta function.

In the following we review the generation of some of the most important non-classical resources in CV quantum information processing, namely the squeezed state, the entangled state and the single photon state. 
Although squeezed states of light have been generated for more than 20 years, significant progress in the production of pure, efficient and stable squeezing has been achieved only recently. Generation of highly squeezed states utilizing the second order nonlinearity via optical parametric amplification and the third order nonlinearity in optical fibers and rubidium vapor have been reported. We will now briefly discuss these approaches.

\subsection{Optical parametric amplification}
\begin{figure}
    \includegraphics[width=9cm]{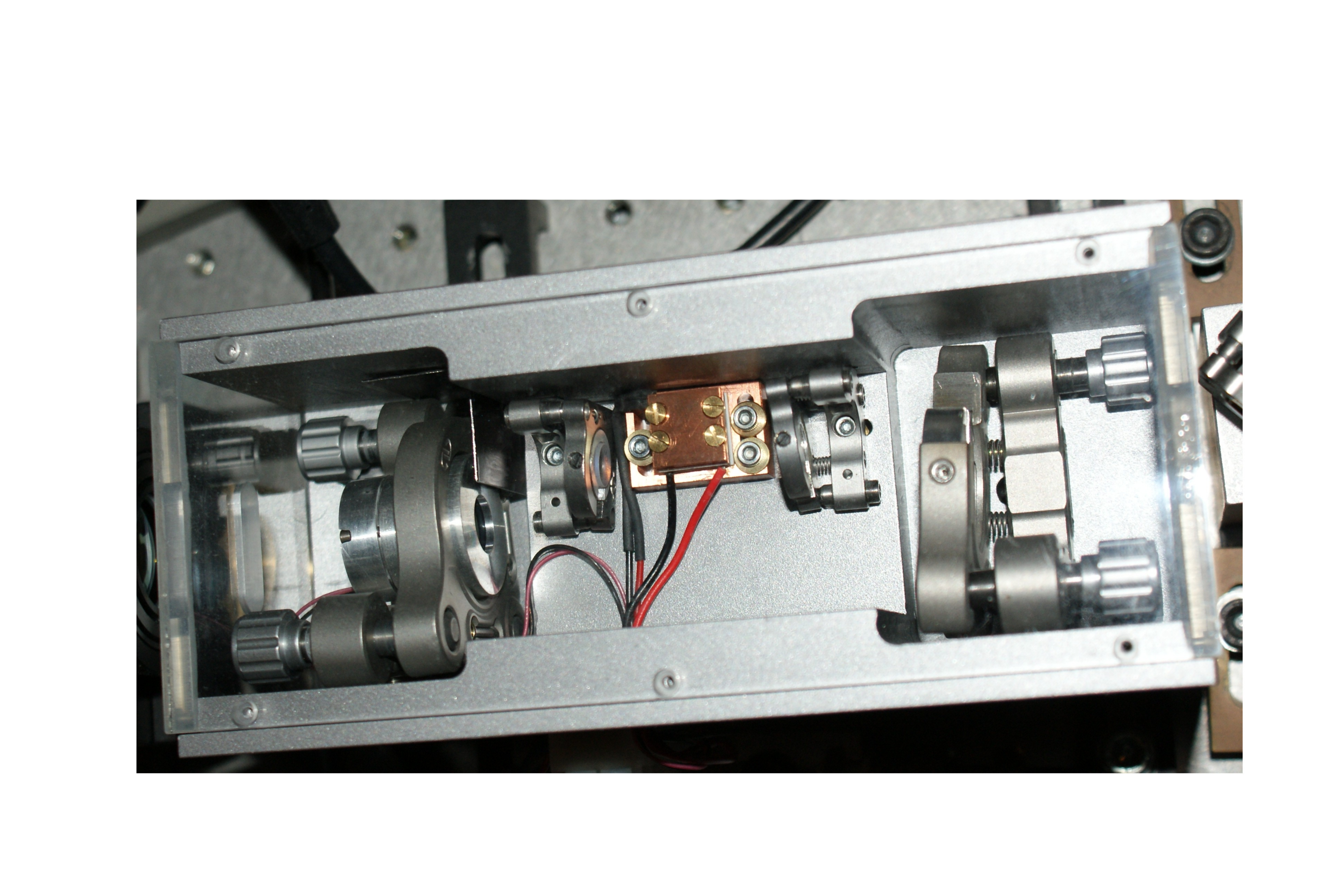}
  \caption{Picture of a bow-tie cavity with a PPKTP crystal used for squeezing the light field. The crystal is placed inside an oven (made of copper) which controls the cystal temperature to ensure optimal phase matching between the waves.}
  \label{OPOsqz}
  \end{figure}

In optical parametric amplification, a pump photon, with frequency $\omega_p$, is injected into a non-linear medium with a second order nonlinearity and breaks up into two new photons; a signal photon of frequency $\omega_s$ and an idler photon of frequency $\omega_i$ such that $\omega_p=\omega_s+\omega_i$~\cite{Louisell1961,Harris1967,Magda1967,Akhnanov1967}.
%%%cite Mandel und aus Russland?? Klyshko??
The two photons are quantum correlated in many degrees of freedom such as time of arrival, the frequency and amplitude and phase quadratures. If the two photons are indistinguishable, meaning that they are in the same spatio-temporal and polarization mode, the quadrature correlations lead to quadrature squeezing. If on the other hand the two modes are physically separable, entanglement between the modes exists in all the degrees of freedom mentioned above. Photon entanglement is in principle maximal for any pump strength whereas the degree of quadrature entanglement is highly dependent on the effective nonlinearity and purity of the generation process. To reach a high effective nonlinearity, different efficiency-enhancing approaches have been applied. One such approach is to apply very short optical pulses with high peak power as a pump field~\cite{Kumar} and another one is to use optical waveguides which prevents the optical pump field from diffracting inside the nonlinear crystal thus increasing the effective nonlinearity~\cite{Serkland1995}. 
%%%Zitat falsch??? wirklich waveguide???
However, the most successful technique demonstrated so far has been to place the nonlinear crystal inside a high quality optical cavity which is resonant for the down-converted squeezed field. First squeezed light generation of this kind was achieved by Kimble and co-workers~\cite{Wu1986.prl}. Already at that time one refered to these squeezed light fields as two mode squeezing~\cite{Milburn1984} as the heterodyne signal detected at a non zero frequency always contains contributions from two entangled light modes, the lower and the upper frequency side band. Operating the optical parametric oscillator above threshold and away from degeneracy with type II phase matching produces two light fields of orthogonal polarization which can be measured separately. This was demonstrated in the pioneering experiment by Giacobino, Fabre, Heidman, Reynaud and
co-workers~\cite{Heidmann1987a,Heidmann1987b} which set the world record of 8.7 dB twin-beam squeezing lasting for a long time. In such a set-up the squeezing is conditional~\cite{Laurat2003}. One of the the two beams has a variance below shot noise conditioned on the measurement of the other beam. More recent experiments use periodic poling to access larger crystal non linearities. 
%%% in fig 10 number is missing, als reference missing
Fig.~\ref{OPOsqz} shows a picture of a bow-tie shaped cavity with a periodically poled KTP crystal. By exploiting a system similar to the one shown in Fig.~\ref{OPOsqz}, as much as 9dB of quadrature squeezing has recently been stably measured in the group of A. Furusawa~\cite{Takeno2007.ox}. The record squeezing to date was however achieved in the group of R. Schnabel who has generated and directly measured more than 10dB squeezing~\cite{Vahlbruch2008.prl}. In this experiment, the non-linear crystal (LiNbO$_3$) constituted the monolithic cavity with mirrors coated onto the curved end facets of the crystal. This was done to minimize the optical losses and improve the phase stability of the system. 
The first experiment using a non-linear monolithic cavity for squeezed light generation  was concerned with frequency doubling, the inverse of parametric amplification, studying in particular the squeezing in the second harmonic mode~\cite{Sizmann1990}.

\subsection{Fiber system}

Another approach for generating squeezed light that has enjoyed significant progress recently is the Kerr effect in fibers~\cite{Shelby1986.prl,Rosenbluh1991.prl,Bergman1991.ol,Dong2008.ol}. As opposed to the three photon mixing process mentioned above, the Kerr effect is a four photon process where two degenerate photons are converted into two new photons (signal and idler). The two converted photons are quantum correlated in many degrees; the quadrature correlations lead to quadrature squeezing as for the optical parametric amplifier mentioned above. Although the two processes (the second-order and third-order processes) are largely equivalent for nearly all realizations, they will differ when very large effective non-linearities are involved. In this regime, the pump beam can no longer be treated as a classical field, and the formation of non-trival quantum states is in principle possible. With present day fiber technology, however, this regime is unattainable in practice. 

Due to the small size of the nonlinear susceptibility for silica, it is common to use long fibers and short pump pulses to effectively increase the Kerr non-linearity and thereby producing an appreciable amount of squeezing. In a very recent experiment a 13.2 m long polarization maintaining fiber was pumped with a 140fs pulse to produce 6.8 dB quadrature squeezing~\cite{Dong2008.ol}. The degree of squeezing was theoretically predicted using a model that among other effects takes into account nonlinear and stochastic Raman effects. It was found that Raman scattering in the fiber markedly deteriorates squeezing at higher energies while guide acoustic wave Brillouin scattering~\cite{Shelby1985.prl} affects the squeezing at lower energies~\cite{Corney2006.prl,Corney2008.pra}.

\subsection{Atomic system}

\begin{figure}
    \includegraphics[width=0.8\columnwidth]{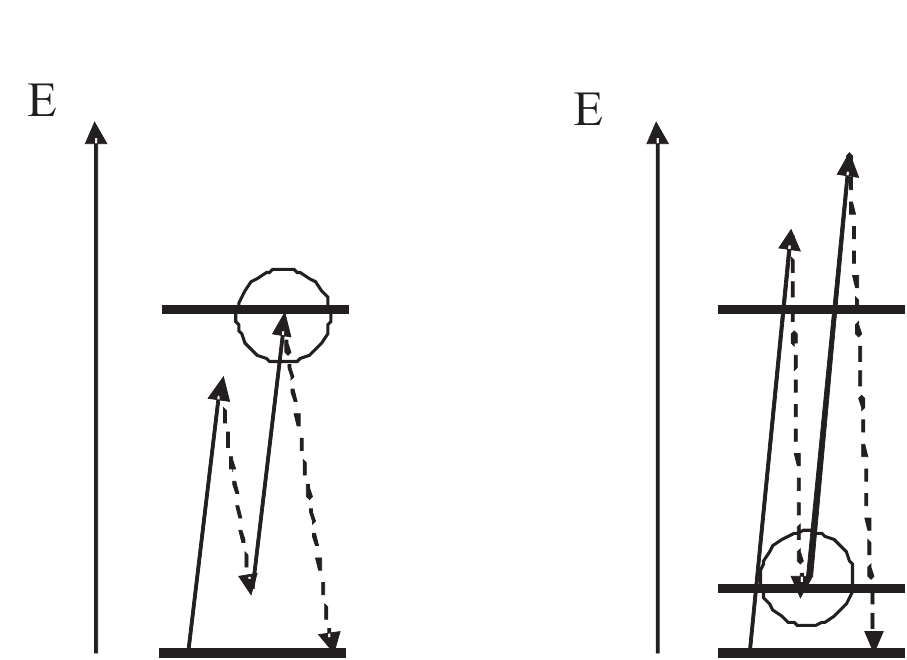}
  \caption{Two atomic level schemes for squeezed light generation.}
  \label{atom}
  \end{figure}

In the first observation of squeezed light in 1985 R.E. Slusher and co-workers used four wave mixing  in a room temperature sodium vapour inside an optical resonator~\cite{Slusher1985.prl}. Four wave mixing is another term for the optical Kerr effect and it produces entangled spectral sidebands symmetrically positioned around the wavelength of the pump light as mentioned above. In there joint interference with the pump radiation the entangled sidebands lead to squeezing~\cite{Huntington2005.pra}. The amount of quantum noise reduction observed by Slusher et al. was 0.3 dB. In the following years several groups tried to improve upon the amount of squeezing achievable in atomic vapours but there was no major advance and the alternative noise reduction using non resonant $\chi^{(2)}$ materials was much more successful. An improvement came about only when E. Giacobino suggested to place laser cooled atom clouds inside a resonator. This resulted in observed 2.2 dB noise reduction~\cite{Lambrecht1996.el}. For more than a decade this record survived. These experiments were operated close to the atomic resonance to enhance the four wave mixing cross section (see Fig.~\ref{atom}a) and it was generally believed that spontaneous emission processes unavoidably limit the achievable squeezing. Only recently the interest in atomic vapour squeezing revived because of the need for narrowband squeezed light. The new attempt took guidance from electromagnetically induced transparency~\cite{Lukin1998.prl} with a detuning far off any excited state resonance at room temperature, with nearly co-propagating pump and probe beams and by making use of ground state coherence (see Fig.~\ref{atom}b). With such a set-up P.D. Lett and co-workers demonstrated 3.5 dB of observed squeezing~\cite{McCormick2007.ol} with much potential for further improvement. This pioneering experiment is likely to mark the beginning of a renaissance of narrow band atomic vapour squeezed light generation. The nearly but not completely co-propagating beams are reminiscent of the dividable entangled beams observed in non degenerate optical parametric oscillators (sec. 5.1). Squeezing is indeed intimately connected to entanglement as it always requires two entangled frequency bands symmetrically positioned on the frequency axis below and above the local oscillator, the carrier or the pump, whatever applies.

\subsection{Single photons}
Optical single photon states can be also generated using the processes presented above: Due to the photon number correlation between the signal and idler fields, a single photon state can be prepared in, say, the signal mode by the measurement of a single photon in the idler mode using a single photon click detector as already described in sec. 4.2. 
%In this case signal and idler must be non-degenerate either in direction, polarization or frequency such that they can be physically separated and one of them can be detected. 
Single photon generation exploiting this heralding method was first accomplished by Mandel et al \cite{Mandel1986}, but in the context of CV using homodyning it was first demonstrated in the experiment of Lvovsky et al \cite{lvovsky2001.prl}. More recently single photon generation in parametric processes has been achieved by several groups and the generated photons have been characterized using either homodyne tomography or single photon counting detectors. The drawback of using parametric processes to generate single photons is that they are probabilistic; due to the intrinsic randomness of the spontaneous generation of pairs of photons the preparation of single photons occurs random in time. There are however many other approaches to the generation of single photons such as the controlled emission of a photon from a quantum dot, from an NV center in diamond and from an atom or an ion placed in a high finesse cavity~\cite{Lounis2005}. 

%Recently tremendous effort has been put on the generation of a superposition of two coherent states which is also referred to as the optical cat state. Such states could be used as logical coherent state qubits. 

\section{Quantum protocols}

\subsection{Quantum key distribution}

Quantum key distribution (QKD) protocols allow two communication parties, Alice and Bob, to generate a shared random key for data encryption with unconditional security. While it can be proved that for classical information transfer there is no means to ensure that no adversary, Eve, has eavesdropped the key distribution, in quantum communication the nature of quantum information itself can be employed to accomplish this goal. 

For CV systems the security of the key exchange relies on measurements of two or more non orthogonal states, e.g. two conjugate quadratures. Due to the uncertainty principle it is never possible for a potential eavesdropper to ascertain simultaneously the exact values of, e.g., the conjugate $X$ and $Y$ quadratures of a displacement in phase space. Thus detecting different quadratures in CV QKD scheme corresponds to measuring qubit states in two different non-orthogonal basis sets, i.e. if a measurement is carried out in one quadrature  $X$ (basis 1) the encoded information of the conjugate quadrature $Y$ (basis 2) is  erased and errors are introduced if a third party tries to listen to the quantum information transfer. The statement can be made even more general: if the information is encoded in two non orthogonal states an eavesdropper can never determine the prepared state with certainty.

In comparison to standard qubit based schemes CV QKD offers several potential advantages: Firstly, the "natural" information carrier of CV communication are coherent quantum states, which can be easily prepared by the use of standard laser beams. In fact, it was a milestone in CV QKD when it was realized that coherent states are sufficient to ensure the security of the protocols and, no more intricate states with distinct "non-classical" features, such as squeezed states or entangled states, are needed \cite{Grosshans2002}. Secondly, the information read-out at the receiver station of Bob is based on homodyne measurements instead of single-photon detection. Thus higher signal repetition rates become feasible and each sent signal pulse actually yields one measurement result. Thirdly, the information content per signal state is not restricted by a binary alphabet, but higher dimensional encoding with a larger number of distinguishable states can be easily accomplished. This last point has led to a reverse development for single photon QKD: novel protocols have been recently designed, which use the CV degrees of freedom of single photons to reach higher bit rates per signal \cite{OSullivan-Hale2005,Walborn2006,Zhang2008}.

For all practical QKD implementations it is important to analyze how experimental imperfections may compromise the security. For CV systems it was believed for a long time that the maximum distance of secure communication was strictly limited to a boundary of less than 50 \% losses. The argument behind this assumption was based on the fact that if more than half of the signal is lost, Eve will possess  a priori information about the signal state, which is higher than the information Bob receives. However, this apparent 3 dB loss limit can be overcome by adapting classical data processing of the CV QKD protocols after state transmission and detection. The method of post-selection puts a threshold on the accepted range of Bob's data, which effectively allow Alice and Bob to select only such quantum measurement results, which produced favorable values for them \cite{silberhorn2002.prl}. As an alternative reverse reconciliation was established for CV-QKD, where the key is drawn from Bob's data instead of Alice's encoded signal states \cite{grosshans2003.nat}. In this protocol the classical information flow is restricted to one way, i.e. only Bob is allowed to send messages to Alice to correct for errors and to perform privacy amplification. This restriction provides an informational advantage to Alice and Bob for all transmission ratios. Notice that the concept of reverse reconciliation has been recently extended to a completely general scenario via the introduction of the reverse secret-key capacity~\cite{Pirandola2009.prl}. Finally, an alternative solution for improving the security thresholds of CV QKD protocols has been identified in the use of multiple quantum communication which leads to an enhancement of security in both direct and reverse reconciliation~\cite{Pirandola2008.natphy}. As a second imperfection, CV QKD can be affected by noise contributions caused by the transmission channel or electronic noise of the homodyne detectors. 

To address the security of a given QKD system and its vulnerability to excess noise, pre-specified groups of eavesdropping attacks are usually considered. There are three different groups of attacks with increasing level of sophistication:
\begin{itemize}
\item{{\it Individual attack:} Eve couples a probe state to the conveyed quantum state, and stores subsequently the probe in a quantum memory. When the measurement basis of Bob has been revealed, Eve extracts the probe states form the memory and measures them one by one.}
\item{{\it Collective attack:} Eve has an ensemble of probe states that interacts (one by one) with the sent states and keeps them in a large quantum memory. After classical authentication, all states in the memory are measured jointly in an optimized measurement using a quantum computer.}
\item{{\it Coherent attack:} Eve has a multi-mode entangled state which is coupled jointly to all the states sent from Alice to Bob, and stored in a large quantum memory. Bob reveals information about the measurement basis, and Eve extracts maximal information from the multi-mode entangled state.}   
\end{itemize}
The security against these attacks has been addressed in a number of publications~\cite{Ralph1999.pra,Hillery2000.pra,Gottesman2001.pra,Reid2000.pra,Cerf2001.pra,ralph2000.pra,silberhorn2002.prl,grosshans2003.nat,weedbroock2004.prl,namiki2004.prl,grosshans2004.prl,grosshans2005.prl,iblisdir2004.prl,navascues2006.prl,patron2006.prl,filip2008.praQKD,heid2008.pra,Pirandola2008.prl,renner2009.prl,patron2009.prl}. We will now briefly discuss some of the most important theory results. 

It has been shown for qubit-based QKD protocols that studying collective attacks is sufficient to ensure unconditional security, and thus the seemingly stronger coherent attack is not more powerful than the collective attack~\cite{renner2005.pra}. These findings have recently been partially extended to continuous variables~\cite{renner2009.prl}, and thus it probably sufficient to study colelctive attacks in CV systems to address full security. Furthermore, it has been proven that among all collective attacks the Gaussian is the optimal one when the alphabet of coherent states is Gaussian~\cite{navascues2006.prl,patron2006.prl}. Inspired by the work of Holevo on the classification of one-mode Gaussian channels~\cite{Holevo2007}, a complete characterization of all collective Gaussian attacks have been recently carried out~\cite{Pirandola2008.prl}. This characterization has further led to a security analysis of the non-switching protocol~\cite{weedbroock2004.prl} (see below) against the most general form of collective Gaussian attacks~\cite{Pirandola2008.prl}. Finally we note that the optimal individual attack was experimentally demonstrated in ref.~\cite{andersen2006.prl,sabuncu2007.pra}  

CV QKD was first experimentally demonstrated by Hirano et al.~\cite{hirano2000.xxx} and by Grosshans et al.~\cite{grosshans2003.nat}. In these experiments the information was encoded either directly into four different coherent states (much like a BB84-type encoding strategy) or continuously into a Gaussian distribution of coherent states. In the latter experiment, Alice randomly varies the amplitude and phase that take on continuous values defined by a Gaussian distribution while Bob randomly measures the amplitude and phase quadrature using homodyne detection. Employing the reverse reconciliation algorithm, Bob then converts the continuous data set into a secret key consisting of binary numbers. Following these proof-of-principle experiments, there has been a number of other CVQKD implementations. E.g. Lorenz et al.~\cite{lorenz2004.apl} used a BB84-type strategy followed by data post-selection as in the experiment of Hirano et al.~\cite{hirano2000.xxx}, but instead of employing the quadrature amplitudes, the Stokes parameters were used as encoding variables. This experiment was recently simplified by considering a two-state protocol where information was encoded into two coherent states. Lance et al.~\cite{lance2005.prl} have implemented a protocol where information is encoded into a Gaussian alphabet as in the experiment of Grosshans et al.~\cite{grosshans2003.nat}, but where Bob performs simultaneous measurements of the conjugate quadrature amplitudes (known as the non-switching protocol); it has been realised that random switching of the measurement basis is not needed to ensure secure key generation~\cite{weedbroock2004.prl,Pirandola2008.prl}. Recently, there has been much progress towards real field implementations of CVQKD: In a recent experiment of Lodewyck et al.~\cite{lodewyck2007.pra}, a reverse-reconciliated QKD system with a secret key rate of 2 kb/s was demonstrated over 25 km of optical fiber. The feasibility of a polarization based QKD system has also  been tested in an experiment where polarization encoded coherent states were transmitted over 100 meter in a free space channel under real atmospheric conditions~\cite{elser2008.xxx}.

\subsection{Distribution of quantum information}

One of the main tasks in quantum information science is to transmit quantum information fault-tolerantly between different nodes in a quantum network. At these nodes, quantum states are prepared stored and processed, and the nodes are linked via quantum channels through which the quantum information is transmitted. The transmission of the quantum states must be done with absolute care due to the fragility of CV quantum information. Direct transmission through quantum channels such as optical fibers and free space channels works for short distances but for large distances such channels introduce loss and noise and the state will eventually decohere into a classical state.  

There are basically two main approaches to circumvent noise and thus errors in the transmission of quantum states between computational units. One method is to employ the idea of quantum teleportation~\cite{bennett1993.prl} combined with entanglement distillation~\cite{bennett1996.prl}, and the other strategy is to use the protocol of quantum error correction coding~\cite{shor1995.pra,steane1996.prl}. These communication protocols have been vastly investigated for discrete variables whereas for CVs, the progress has been somewhat slow. Recently, however, new ideas and techniques have appeared making some of these protocols feasible under certain assumptions. 

\subsubsection{Teleportation and distillation} \label{sec:entdist}
Quantum teleportation is the protocol of transmitting quantum information by the joint action of a perfect classical channel and a quantum channel over which entanglement is shared between the sender and the receiver. The main challenge therefore reduces to the distribution of entanglement through noisy quantum channels. CV teleportation was first suggested by Vaidman in 1994 \cite{Vaidman1994} (based on the work of Aharonov and Albert~\cite{Aharanov1,Aharanov2}) and transferred to real photonic systems in 1998\cite{Braunstein1998,Ralph1998}. It was experimentally realized in 1998 with a fidelity between the input and output state of F=58\%~\cite{furusawa1998.sci}, and since then refined experimental methods have pushed the fidelity to values as high as F=83\%~\cite{yukawa2008.pra}. Moreover teleportation of a single mode squeezed state and entangled states have been demonstrated~\cite{yonezawa2007.prl,takei2005.prl} as well as teleportation in a network~\cite{yonezawa2004.nat,yonezawa2007.pra}. The fidelity for which quantum teleportation has been carried was theoretically analysed in ref.~\cite{hammerer2005.prl,calsamiglia2008.xxx,owari2008.njp}. 

It is also interesting to note that the teleportation protocol can be envisaged as an operation that implements the identity transformation. It is however possible to modify the entangled state so as to enforce a desired operation of the input state~\cite{Bartlett2003.pra}. The advantage of such a teleportation based operation is that the difficult transformation is applied off-line to the entanglement resource, whereas the transformation of the quantum information is carried out through a deterministic and clean teleportation operation. 

As mentioned above, for teleportation to work efficiently, entanglement must be distributed faithfully between the nodes in the quantum network. However, the entanglement distribution usually takes place in noisy quantum channels and thus the conveyed quantum states are corrupted and the states must be cleaned up. This can be done by means of a probabilistic quantum distillation protocol which distills  a small ensemble of highly entangled states out of a larger ensemble of less entangled states~\cite{bennett1996.prl}. It has been shown theoretically that if the corrupted entangled states are Gaussian (which is for example the case for Gaussian entanglement that has undergone a constant loss), the distillation protocol cannot be implemented with Gaussian transformations~\cite{eisert2002.prl,fiurasek2002.prl,giedke2002.pra}; one must resort to the difficult non-Gaussian transformations. Although a number of schemes for entanglement distillation of Gaussian states have been devised~\cite{duan2000.prl,browne2003.pra,fiurasek2003.pra} there has yet been no full experimental demonstration~\cite{ourjoumtsev2007.prl}, which could counteract decoherence introduced by constant losses.  
On the other hand, if the noisy channel introduces non-Gaussian noise as opposed to Gaussian noise, distillation can be carried out relatively easily as demonstrated in  recent experiments. For example, in the experiment by Dong et al.~\cite{dong2008.nat} entanglement was transmitted through a channel with a time varying transmission coefficient and in the experiment by Hage et al.~\cite{hage2008.nat} the channel introduced phase noise to the state. In both cases the resulting state was a mixed non-Gaussian state that could be distilled using linear optics, homodyne detection and feed-forward.

\subsubsection{Quantum error correction coding}
Classical error correction coding is widely used to battle errors in classical devices. The main idea of all classical correction coding systems is to introduce redundancy in the encoded information so that even though some bits will be corrupted, a larger majority will be untouched and thus, much like a majority vote, the original information can be recovered through direct measurements and recreation. 

Just as classical error correction coding is enabling efficient classical processing, quantum error correction coding (QECC) is believed to be one of the technologies that eventually may allow for fault-tolerant quantum information processing. However, the classical encoding and decoding strategy would not work on quantum information because of  the no-cloning theorem and the fact that a measurement inevitably disturbs the measured state. The conceptual idea of QECC is to encode the information as a subsystem of a larger system. This results in a multi-mode entangled state in which the information is embedded. The type of the entangled state is determined by the error that it should protect the information against. After transmission in the noisy environment, the information is decoded using partial detection and feed-forward: The error that might have occurred is diagnosed through a so-called syndrome measurement and the outcome is used to make a corrective transformation~\cite{shor1995.pra,steane1996.prl}. First suggestions to extend the scheme to continuous variables were put forward by Braunstein~\cite{braunstein_error_1998} and Lloyd and Slotine~\cite{lloyd_slotine_1998}, and a linear circuit implementing CV QECC was first suggested by Braunstein~\cite{braunstein1998.nat}. He presented a scheme that encoded the quantum information into an 8-mode squeezed state which was dispersed into nine different quantum channels. After transmission, the nine modes interfere in a beam splitting array and a syndrome measurement is carried out on eight modes using homodyne detection. The error is then spotted and subsequently corrected in the remaining mode. Any errors imposed onto a single channel can be perfectly and deterministically corrected under the assumption that infinitely squeezed input states are used for the encoding and the added noise is non-Gaussian. The QECC protocol has very recently been implemented~\cite{aoki2008.xxx}. Another code for CV quantum error correction based on distributed entanglement has been devised by Wilde et al.~\cite{wilde2007.pra}. 

A quantum code for protecting coherent states from complete erasure noise has recently been suggested~\cite{niset2008.prl}. The quantum error erasure protocol allows for the faithful transmission of coherent states through channels which either erase the information or perfectly transmit it. The scheme encodes two coherent states into a bi-partite entangled state through linear interference, and the resulting four mode entangled state is conveyed through four independent channels. The transmitted state is corrected by reversing the linear interferences, performing a syndrome measurement and finally executing a corrective transformation (corresponding to linear displacement of the remaining modes). This transformation, however, depends on the location of the error which might not be known to the receiver. The corrective action can however be independent of the location of the error by substituting the deterministic displacement operation with a probabilistic heralding operation~\cite{niset2008.prl}. An experiment demonstrating CV error erasure coding (deterministically and probabilistically) has recently been implemented~\cite{Lassen2009}. We also note that another probabilistic error erasure correction scheme was recently implemented by Wittmann et al.~\cite{wittmann2008.pra};  it is, however, not displacement preserving.

\subsection{Quantum memory}
Another very important ingredient in quantum information processors is the quantum memory which is capable of storing quantum information faithfully. For many quantum informational routines the quantum memory element is a crucial tool; e.g. the optimal eavesdropping attack relies on a quantum memory, a quantum repeater station is based on a quantum memory and scalable quantum computing requires a quantum memory. Due to all these potential applications of the quantum memory there has recently been substantial effort in constructing efficient memories for optical quantum states.

For storing CV quantum information, basically three different approaches have be applied~\cite{Hammerer2008.xxx}. They are all based on the efficient coupling between a light field and a large collection of atoms - an atomic ensemble, but the type of interaction differ among the three approaches. One approach is based on a QND interaction and feedback, another relies on Electro-magnetically Induced Transparency (EIT) via Raman interaction and the last one is based on photon echo techniques.    

The first realization of a quantum memory (outperforming a classical memory) was based on a QND type operation between the light and the atomic ensemble followed by an electro-optical feedback system~\cite{Julsgaard2004.nat}. In this experiment a pulse of light carrying quantum information was stored in two cesium ensembles with counter-rotating spins and efficiently retrieved to yield an output quantum state with a better quality than could be achieved by any classical memory. 

The second realization of a coherent memory was based on a EIT. EIT is the effect of making the atomic ensemble transparent for a signal pulse by applying a strong control pulse~\cite{Lukin2003}. In addition to the creation of transparency, the group velocity is dramatically increased which has the effect of slowing down the pulse. If the control pulse is switched off, the signal pulse stops and creates a so-called dark state polariton wave. By turning on the control pulse, the process is coherently reversed and the initial pulse is retrieved. This approach has been used in several recent experiments on storing quantum states described by discrete variables~\cite{Chaneliere2005,Choi2008,Eisaman2005} as well as quantum states characterised by continuous variables. In these latter experiments, squeezed vacuum states were stored in clouds of Rubidium and the total storage efficiency was about 10-15\% \cite{Appel2008.prl,Honda2008.prl,Hetet2008}, and an experiment in Cesium vapor demonstrated the storage and retrieval of a quantum state of light without adding noise in the process~\cite{Cviklinski2008}. 

The third approach to implement a quantum memory is based on a photon echo technique~\cite{Moiseev2001.prl,Kraus2006.pra}. In this memory the light pulse is stored through absorption in a medium which is intentionally inhomogenously broadened to enhance the bandwidth. The broadening causes dephasing which is then compensated by using the photon echo technique. In addition, the photon echo also allows to control the release of the stored quantum state. This approach has been experimentally realized using solid state media consisting of Pr3$^+$ ions in a Y$_2$SiO$_5$ host and Er3$^+$ in LiNbO$_3$ ~\cite{Alexander2006.prl,Staudt2007.prl,Hetet2008.prl}.

\subsection{Quantum computation}
\label{computing}
Quantum computing can be carried out by encoding quantum information in states of quantum systems and subsequently executing a set of universal unitary operations, also called quantum gates~\cite{nielsen2000.book}. Enormous progress has been made in qubit-based quantum computing, both on a theoretical and on an experimental level. On the contrary,  research in CV quantum computing has started later and has not progressed as far. This is also  due to the difficulties faced in the experimental realization of the required large non-linearities: It has been shown that universal CV quantum computing can be only performed by the use of a strong non-linearity leading to a non-Gaussian transformation~\cite{lloyd1999.prl}. In fact, by utilizing only Gaussian transformations, the resulting computational quantum circuit will have no advantage over classical computers~\cite{bartlett2002.prl}. Therefore, the holy grail in building a CV quantum computer is thus the efficient implementation of a non-Gaussian transformation. 

There are basically two circuit models for realizing CV quantum computation. The conventional approach where the difficult computational gates are implemented directly onto the quantum information, and the off-line based approach where the difficult non-Gaussian operations are moved off the computational line. In the conventional scheme, the fragile quantum information is launched directly into a non-linear medium thus involving some non-linear coupling~\cite{lloyd1999.prl}. This approach is experimentally very challenging due to the difficulty in implementing pure, efficient and deterministic non-Gaussian quantum gates. It was therefore an important discovery when Knill, Laflamme and Milburn~\cite{knill2001.nat} found that the difficult operations can be placed off-line whereas the interactions involving the quantum information are carried out using solely simple linear optics and photon counting. This is a very useful approach since the difficult off-line operations can be carried out probabilistically without corrupting the in-line computation. A similar result has been found for CV: By preparing a special non-Gaussian state off-line (possibly probabilistically) a perfect gate operation can be performed deterministically onto a CV state using only Gaussian transformations, homodyne detection and feed-forward~\cite{gottesman2001.pra,bartlett2002.pra,ralph2003.pra,ghose2007.jmo}. This will, of course, require a reliable high performance quantum memory. We should further note that the information encoding for such schemes is rather complex, since it relies on the hybrid encoding of qubits in harmonic oscillators, either as squeezed states~\cite{gottesman2001.pra} or coherent states~\cite{ralph2003.pra}. 

Another CV computer, based on off-line prepared resources, is the co-called cluster state computer~\cite{menicucci2006.prl,loock2007.josa} following the initial proposals for discrete variables \cite{Raussendorf2001}. In this approach the off-line prepared resources are Gaussian multi-mode entangled states (cluster states) which are relatively easy to prepare (compared to non-Gaussian resources). As in the previous approach, once the resource has been prepared, the gate operation can be implemented using only linear optics, detection and feedforward. Using homodyne detection an arbitrary Gaussian transformation can be implemented, which however is insufficient for universal computation. Universality for the cluster state computer can be introduced by using a non-Gaussian measurement in replacement of the homodyne detector. Therefore, by using e.g. a photon counting device (which is non-Gaussian), universal computation is possible with Gaussian state resources~\cite{menicucci2006.prl}. 

The three approaches to CV quantum computation mentioned above are basically differentiated by the position of the non-Gaussian transformation; it is either placed in-line with the computation, it is placed off-line and used to prepare non-Gaussian resources or it is placed at the measurement stage.

\section{Conclusion}

One and the same quantum system can be described with either discrete or
continuous quantum variables. However, depending on the situation one of the
two descriptions can be much more efficient than the other. In the last 20
years main stream quantum information research developed concentrated on the
use of discrete variable descriptions. In this review we have focussed on
the alternative: quantum information processing with continuous variables.

We have been concluding this review on continuous variable quantum information processing with the most ambitious and futuristic application of quantum continuous variables - quantum computing. While the protocol of quantum key distribution is reaching the level of commercialization, many other protocols are still in its infancy. The main reason for this is the difficulty in enabling a controlled non-Gaussian operation which is required for carrying out universal quantum operations on continuous variables. The complete mastering of such a non-Gaussian operation is still to come and will, when it comes, probably create a breakthrough in the field of CV quantum information. Despite the lack of an efficient and controllable non-Gaussian operation, it is still possible to carry out the protocols of teleportation, quantum key distribution, certain classes of entanglement distillation and quantum error correction coding protocols and quantum memory. All these applications have been reviewed in this paper. We anticipate that more protocols will be invented and more applications will be found because the field is still progressing at a fast pace.

{\bf Acknowledgments} We gratefully acknowledge support from the EU (project COMPAS), the Danish Research Council and the Deutsche Forschungsgemeinschaft.

%%% Give your bibliography:

\end{document}